\documentclass[11pt]{article}

\usepackage{amsmath}
\usepackage{url}
\usepackage{xspace}

\newcommand{\code}[1]{\texttt{#1}}
\newcommand{\T}{\ensuremath{{\cal T}}\xspace}
\newcommand{\vars}{\ensuremath{X}}
\newcommand{\mktuple}[1]{\ensuremath{\langle #1\rangle}}
\newcommand{\eqbydef}{\ensuremath{\:\dot{=}\:}}

\begin{document}

\title{\textbf{The VMT-LIB Language and Tools}}
\author{Alessandro Cimatti, Alberto Griggio, Stefano Tonetta}%
\renewcommand{\today}{Fondazione Bruno Kessler}
\maketitle

\begin{abstract}
  We present VMT-LIB, a language 
  for the representation of verification problems of linear-time temporal properties 
  on infinite-state symbolic transition systems.
  VMT-LIB is an extension of the standard SMT-LIB language for SMT solvers,
  developed with the goal of facilitating the interoperability and exchange of benchmark problems among different verification tools. 
  Besides describing its syntax and semantics, we also present a set of open-source tools to work with the language.
\end{abstract}

\section{Introduction}

In this paper we describe VMT-LIB, 
a language for the representation of verification problems 
of linear-time temporal properties on infinite-state symbolic transition systems.
VMT-LIB was designed with the main goal of being simple to use (i.e. to parse and generate)
for verification tools, 
with the aim of facilitating the interoperability among different tools and the collection of verification benchmarks for infinite-state systems.
As such, it is similar in spirit to the Aiger~\cite{aiger} language for finite-state systems,
and to the BTOR~\cite{btor} language for word-level systems with arrays,
with the difference that VMT-LIB supports arbitrary background SMT theories,
including e.g. linear and nonlinear arithmetic, uninterpreted functions, and quantifiers.

VMT-LIB was developed as an extension 
of the standard SMT-LIB~\cite{smtlib2} language for SMT solvers,
by exploiting the capability of SMT-LIB of attaching metadata to terms and formulas via annotations.
In particular, a valid VMT-LIB file is also a valid SMT-LIB file. 
This allows to reuse all the libraries for manipulating SMT-LIB formulas that are available for various languages (e.g.~\cite{pysmt,smtswitch,javasmt}).
Besides these generic libraries, we have also developed a set of tools to work with the language, including converters to and from other formats and formalisms (including Aiger, BTOR and Constrained Horn Clauses).
All the tools are open source and available at the VMT-LIB webpage~\cite{vmtlib}.
VMT-LIB is fully supported by the nuXmv~\cite{nuxmv} model checker,
and it has been used as a benchmark format in several publications over the last few years (e.g.~\cite{ic3ia,prophic3,avr,lambda,incrlin}).

The rest of the report is structured as follows. 
After providing the necessary theoretical background in \S\ref{sec:background},
we describe the VMT-LIB syntax in \S\ref{sec:syntax}, and its semantics in \S\ref{sec:semantics}.
In \S\ref{sec:tools} we describe a set of open-source tools that we have developed to work with the language.
Finally, we conclude in \S\ref{sec:conclusions}.

\section{Theoretical Background and Definitions}
\label{sec:background}

Our setting is many-sorted first order logic.
We use the standard notions of theory, satisfiability, validity, and logical consequence.
We refer to the SMT-LIB specifications~\cite{smtlib2} for more details.
We denote generic theories as \T. We
write $\varphi \models_\T \psi$ to denote that the formula $\psi$ is a
logical consequence of $\varphi$ in the theory \T; when clear from context,
we omit \T and simply write $\varphi \models \psi$.
%

We refer to 0-arity predicates as Boolean variables, and to 0-arity uninterpreted functions as (theory) variables.
%

Given a set of variables $\vars$, a signature $\Sigma$, a domain $M$,
an interpretation function $\mathcal{I}$ of the symbols in $\Sigma$ on
the domain $M$, an assignment $\sigma$ to the variables in $\vars$ on
the domain $M$, and a $\sigma$-formula $\phi(\vars)$ with free
variables in $\vars$, the satisfaction relation
$\mktuple{M,\mathcal{I}}\models\phi$ is defined in the usual
way. 

%

For each variable $x$, we assume that there exists a corresponding variable $x'$,
called the \emph{primed version} of $x$.
If $\vars$ is a set of variables, $\vars'$ is the set obtained by
replacing each element $x$ with its primed version ($\vars' = \{x'
\mid x \in \vars\}$).
$\varphi'$ is the formula obtained by
replacing each occurrence variable in $\varphi$ with the corresponding primed.
%


In the following, the signature $\Sigma$ and the theory \T
are implicitly given.
A \emph{transition system (TS)} $S$ is a tuple $\mktuple{X,I(X),T(Y,X,X')}$ where
$X$ is a set of \emph{state} variables, $I(X)$ is a formula representing
the initial states, and $T(Y,X,X')$ is a formula representing the
transitions, where $Y$ is a set of \emph{input} variables. 
%

\section{Syntax}
\label{sec:syntax}

VMT-LIB exploits the capability offered by the SMT-LIB language of attaching
metadata to terms and formulas in order to specify the
components of the transition system and the properties to verify.
More specifically, we use the following SMT-LIB \emph{annotations}:

\begin{description}

\item[\texttt{:next name}] is used to represent state
  variables.  For each variable $x$ in the model, the VMT-LIB file
  contains a pair of variables, $x^c$ and $x^n$, representing
  respectively the current and next version of $x$.  The two variables
  are linked by annotating $x^c$ with the attribute \texttt{:next} $x^n$.
  All the variables that are not in relation with another by means
  of a \texttt{:next} attribute are considered inputs.
  Note that \texttt{:next} must define an injective function (i.e. it is an error if there are two variables with the same \texttt{:next} value), 
  and that the names of the variables are not important.

\item[\texttt{:init}] is used to specify the formula for the
  initial states of the model. This formula should contain neither
  next-state variables nor input variables. 
  Multiple formulas annotated with \texttt{:init} are implicitly conjoined.
  As a convenience, the annotation can also use a ``dummy'' value
  \texttt{true}.

\item[\texttt{:trans}] is used to specify the formula for the
  transition relation. As in the case for \texttt{:init}, 
  multiple \texttt{:trans} formulas are conjoined together, 
  and also in this case the annotation can be written as \texttt{:trans true}.

\item[\texttt{:invar-property idx}] is used to specify
  invariant properties, i.e. formulas of the form $G p$, where $p$ is
  the formula annotated with \texttt{:invar-property}. 
  The non-negative integer \emph{idx} is a unique identifier for the
  property.

\item[\texttt{:live-property idx}] is used to specify an LTL
  property of the form $\mathbf{F} \mathbf{G} p$, where $p$ is the formula annotated
  with \texttt{:live-property}. 
  The non-negative integer \emph{idx} is a unique identifier for the property.
\end{description}

In a VMT-LIB file, only annotated terms and their sub-terms are
meaningful. Any other term is ignored.  Moreover, only the following
commands are allowed to occur in VMT-LIB files:
\code{set-logic},
\code{set-option},
\code{declare-sort},
\code{define-sort},
\code{declare-fun},
\code{define-fun}.%
(For convenience, an additional \code{(assert true)} command is
allowed to appear at the end of the file.)

\noindent The following example shows a simple  model in the syntax of nuXmv~\cite{nuxmv} on the left, and
its corresponding VMT-LIB translation on the right. \\ \par

\noindent
\begin{footnotesize}
  \begin{tabular}{|c|c|}
    \hline
    \textbf{nuXmv} & \textbf{VMT} \\
    \hline
    \begin{minipage}{0.32\linewidth}
\begin{verbatim}

MODULE main
-- declaring the state 
-- variable x
VAR x : integer;
IVAR b : boolean;
INIT x = 1;
TRANS 
next(x) = b ? x + 1 : x;
INVARSPEC x > 0;
LTLSPEC FG x > 10;

\end{verbatim}
    \end{minipage}
&
    \begin{minipage}{0.6\linewidth}
\begin{verbatim}

; declaring the state variable x
(declare-const x Int)
(declare-const x.next Int)
(define-fun sv.x () Int (! x :next x.next))

(declare-const b Bool)
(define-fun init () Bool 
         (! (= x 1) :init))
(define-fun trans () Bool
   (! (= x.next (ite b (+ x 1) x)) :trans)
(define-fun p1 () Bool 
              (! (> x 0) :invar-property 1))
(define-fun p2 () Bool 
              (! (> x 10) :live-property 2))

\end{verbatim}
    \end{minipage} \\
    \hline
  \end{tabular}
\end{footnotesize}
\ \\

Since the SMT-LIB format 
(and thus also the VMT-LIB one that inherits from SMT-LIB) 
does not allow to annotate the declaration of variables, it is
a good practice to insert immediately after the declaration of the
variables a set of defines to specify the relations among variables.
See for instance the define \code{sv.x} in the example above that
introduces the relation between \code{x} and \code{x.next}.

\paragraph{LTL Properties, Invariant Constraints and Fairness Conditions.}
Since one of the main goals of VMT-LIB is that of simplicity,
the language does not provide any direct support for high-level constructs
such as specifications written in full LTL, invariant constraints or fairness conditions.
However, this is not a limitation in terms of expressiveness, as all such constructs can be easily encoded in VMT-LIB:

\begin{description}
\item[LTL properties] can be compiled into invariant and/or live properties using standard algorithms from the literature (e.g. \cite{Vardi95,ltl2smv,DBLP:conf/fmcad/ClaessenES13});

\item[invariant constraints] can be straightforwardly embedded into init and trans formulas;

\item[fairness conditions] can be embedded into live properties using a symbolic version of standard degeneralization procedures for B\"uchi automata (e.g. \cite{ltl-proofs}).
\end{description}

\section{Semantics}
\label{sec:semantics}

Given a transition system $S \eqbydef \mktuple{X, I(X), T(Y, X, X')}$ over a background theory $T$ with a signature $\Sigma$ and an interpretation ${\cal I}$,
a \emph{state} $s$ of $S$ is
an interpretation of the state variables $X$.
A (finite) \emph{path} of $S$ is a finite sequence $\pi\eqbydef
s_0,s_1,\ldots,s_k$ of states, with the same domain and interpretation
of symbols in the signature $\Sigma$, such that ${\cal I}, s_0\models I(X)$ and for all $i$,
$0\leq i<k$, ${\cal I}, s_i, s'_{i+1} \models \exists Y. T(Y, X, X')$.  We say that a state $s$ is
reachable in $S$ iff there exists a path of $S$ ending in $s$.
Note that, since the interpretation ${\cal I}$ is unique,
uninterpreted function and predicate symbols are \emph{rigid}, 
i.e. they are not allowed to change across states.

\subsubsection*{Invariant Properties}
An \emph{invariant property} $p$ is a symbolic representation of a set of states that must be a superset of the reachable states of $S$. In other words, $S \models p$ iff $\forall s. s$ is reachable in $S$, $s \models p$.
Consequently, a \emph{counterexample} for $p$ is a \emph{finite path} $s_0, \ldots, s_k$ of $S$ such that $s_k \models \neg p$.

\subsubsection*{Live Properties}
A \emph{live property} $p$ represents a set of states that is \emph{eventually invariant}. In LTL syntax, it would be denoted with $\mathbf{FG}p$.
More formally, $S \models p$ iff for all paths $s_0, \ldots, s_i, \ldots$, $\exists i. \forall j > i. s_j \models p$.
(Note that finite paths $s_0, \ldots, s_k$ vacuously satisfy a live property, because we can always take $i=k$ to satisfy the previous definition.)
Consequently, a \emph{counterexample} for $p$ is an \emph{infinite path} $s_0, \ldots, s_i, \ldots$ of $S$ such that $\forall i. \exists j > i. s_j \models \neg p$.


\section{VMT-LIB Tools}
\label{sec:tools}

\paragraph{VMT-LIB support in verification tools.}
The VMT-LIB language is fully supported by nuXmv~\cite{nuxmv}, 
a state-of-the-art symbolic model checker for finite- and infinite-state systems. 
Recently, the language has been adopted also by the AVR~\cite{avr} model checker. 
VMT-LIB is also the native language of ic3ia~\cite{ic3ia-tool}, 
an efficient open-source model checker for invariant and LTL properties,
as well as its recent extensions ProphIC3~\cite{prophic3} (for discovering universally quantified invariants over arrays) 
and Lambda~\cite{lambda} (for the verification of parametric systems).

\paragraph{Tools for working with VMT-LIB.}

We provide a set of tools (mainly written in the Python programming language) 
to work with the VMT-LIB language. They are all available from the VMT-LIB webpage~\cite{vmtlib}. Currently, the following tools are provided:

\begin{description}
\item{\bf vmt.py:} parsing and printing of transition systems in VMT-LIB.

\item{\bf vmt2btor.py:} converter from VMT-LIB to the BTOR format.

\item{\bf btor2vmt.py:} converter from BTOR to VMT-LIB.

\item{\bf vmt2horn.py:} converter from VMT-LIB to Constrained Horn Clauses.

\item{\bf vmt2nuxmv.py:} converter from VMT-LIB to the SMV dialect of nuXmv.

\item{\bf ltl2vmt.py:} a tool to convert arbitrary LTL properties into VMT-LIB \texttt{:live-property} specifications, by compiling them into symbolic tableaux which are then put in product with the transition system.  
\end{description}

\noindent Moreover, further converters to VMT-LIB are available through nuXmv and ic3ia. In particular, ic3ia provides a \texttt{horn2vmt} tool for converting Constrained Horn Clauses to VMT-LIB,
whereas nuXmv can be used to convert from VMT-LIB to Aiger and vice versa.

\section{Conclusions}
\label{sec:conclusions}

We have presented VMT-LIB, a language and a set of tools for the specification of verification problems over infinite-state transition systems aimed at simplicity and interoperability.
In the future, we plan to extend the format to support the representation of counterexample traces for violated properties, 
and possibly also proof certificates for verified properties.

\bibliographystyle{plain}
\bibliography{main}

\end{document}